\documentstyle[aps,prd,epsfig,preprint]{revtex}
\tightenlines
\def\fun#1#2{\lower3.6pt
\vbox{\baselineskip0pt\lineskip.9pt
\ialign{$\mathsurround=0pt#1\hfill##\hfil$
\crcr#2\crcr\sim\crcr}}}
\newcommand{\newc}{\newcommand} 
\newc{\ra}{\rightarrow} 
\newc{\lra}{\leftrightarrow} 
\newc{\beq}{\begin{equation}} 
\newc{\eeq}{\end{equation}} 
\newc{\barr}{\begin{eqnarray}} 
\newc{\earr}{\end{eqnarray}} 
\begin{document}
\vspace{0.5in}
\title{\vskip-2.5truecm{\hfill \baselineskip 14pt 
{\hfill {{\small \hfill UT-STPD-2/00}}}\\
{{\small \hfill FISIST/12-2000/CFIF}}
\vskip .1truecm} 
\vspace{1.0cm}
\vskip 0.1truecm{\bf Cold Dark Matter detection in SUSY models
at large $\tan\beta$ }}
\vspace{1cm}
\author{{M. E. G\'omez}$^{(1)}$\thanks{mgomez@cfif.ist.utl.pt}
{and J. D. Vergados}$^{(2)}$\thanks{vergados@cc.uoi.gr}} 
\vspace{1.0cm}
\address{$^{(1)}${\it Centro de F\'{\i}sica das 
Interac\c{c}\~{o}es Fundamentais (CFIF),  
Departamento de F\'{\i}sica, \\ Instituto Superior T\'{e}cnico, 
Av. Rovisco Pais, 1049-001 Lisboa, Portugal.}} 
\address{$^{(2)}${\it Theoretical Physics Section, 
University of Ioannina, \\
Ioannina  GR , Greece.}}
\maketitle

\vspace{.8cm}

\begin{abstract}
\baselineskip 12pt
\par

We study the direct detection rate for SUSY cold dark matter (CDM) predicted
by the minimal supersymmetric standard model with universal boundary 
conditions and large values for $tan\beta$.
The relic abundance of the lightest supersymmetric particle (LSP),
assumed to be approximately a bino, is
obtained by including its coannihilations with the next-to-lightest 
supersymmetric particle (NLSP), which is the lightest s-tau. The
cosmological constraint on this quantity severely limits the allowed
SUSY parameter space, especially in the case the $CP$-even 
Higgs has mass of around 114 $GeV$.
We find that for large tan$\beta$
it is possible to find a subsection of the allowed 
parameter space, which yields detectable rates in the currently
 planned experiments.

\end{abstract}

\thispagestyle{empty}
\newpage
\pagestyle{plain}
\setcounter{page}{1}
\baselineskip 20pt
In recent years the consideration of exotic dark matter has become necessary
in order to close the Universe \cite{Jungm}. Furthermore in
in order to understand the large scale structure of the universe 
it has become necessary to consider matter
made up of particles which were 
non-relativistic at the time of freeze out. This is  the cold dark 
matter component (CDM). The COBE data ~\cite{COBE} suggest that CDM
is at least $60\%$. On the other hand during the last few years
evidence has appeared from two different teams,
the High-z Supernova Search Team \cite {HSST} and the
Supernova Cosmology Project  ~\cite {SCP} 
 which suggests that the Universe may be dominated by 
the  cosmological constant $\Lambda$.
As a matter of fact recent data the situation can be adequately
described by  a baryonic component $\Omega_B=0.1$ along with the exotic 
components $\Omega _{CDM}= 0.3$ and $\Omega _{\Lambda}= 0.6$.

In fact the DAMA experiment
has claimed the observation of one signal in direct detection of a WIMP, which
with better statistics has subsequently been interpreted as a modulation signal
~\cite{BERNA1}.

The above developments are in line with particle physics considerations. Thus,
in the currently favored supersymmetric (SUSY)
extensions of the standard model, the most natural WIMP  and candidate 
for CDM is the LSP,
i.e. the lightest supersymmetric particle. In the most favored scenarios the
LSP is the lightest neutralino, which can be simply described as a 
Majorana fermion, a linear 
combination of the neutral components of the gauginos and Higgsinos
\cite{Jungm},\cite{ref1}-\cite{KVprd}. Its stability is guaranteed 
by imposing R-parity conservation , which implies that the LSP's 
can disappear only by annihilating in pairs. Since this particle 
is expected to be very massive, $m_{\tilde{\chi}} \geq 30 GeV$, and
extremely non relativistic with average kinetic energy $T \leq 100 KeV$,
it can be directly detected ~\cite{JDV96}-\cite{KVprd} mainly via the recoiling
of a nucleus (A,Z) in the elastic scattering process:
$\tilde\chi \, +\, (A,Z) \, \to \, \tilde\chi \,  + \, (A,Z)^*$ 

In order to compute the event rate one needs
the following ingredients:

1) An effective Lagrangian at the elementary particle 
(quark) level obtained in the framework of supersymmetry as described 
, e.g., in Refs.~\cite{Jungm}, \cite{Hab-Ka} and 
Bottino {\it et al} \cite{ref2}. 

2)  A procedure in going from the quark to the nucleon level, i.e. a quark 
model for the nucleon. The results depend crucially on the content of the
nucleon in quarks other than u and d. This is particularly true for the scalar
couplings as well as the isoscalar axial coupling ~\cite{Dree}- \cite{Chen}.

3) Compute the relevant nuclear matrix elements, 
using as reliable as possible many body nuclear wave functions.
 A  complete list of references can be found in recent publications
~\cite{JDV00}-\cite{Verg00}.

In the present work we are going to be primarily concerned with the first item
of the above list. We will be concerned, in particular, with SUSY models
predicting a LSP with a relic abundance of cosmological 
interest. Following the considerations of ~\cite {cdm} on the composition 
of the energy density of the universe in scenarios with vanishing 
and non vanishing cosmological constant we assume $\Omega_{LSP} h^2$ 
in the range:

\begin{equation}
0.09 \le \Omega_{LSP} h^2 \le 0.22
\label{eq:in2}
\end{equation}
 
Combining items 1) and 2) we can obtain the the LSP-nucleon cross section. We
will limit ourselves in the experimentally interesting range:
\begin{equation}
4\times 10^{-7}~pb~ \le \sigma^{nucleon}_{scalar} 
\le 2 \times 10^{-5}~pb~
\label{eq:in3}
\end{equation}
We should remember, however, that the event rate does not depend only
on the nucleon cross section, but on other parameters also, mainly
on the LSP mass and the nucleus used in target. 

The calculation of this cross section  has become pretty standard.
One starts with   
representative input in the restricted SUSY parameter space as described in
the literature, e.g. Bottino {\it et al.} ~\cite{ref2}, 
and Arnowitt {\it et al.} \cite {ref4}, \cite {bot}, and recently
\cite {cross}.

In the present work will show how the more restrictive version of the
MSSM, with minimal supergravity (mSUGRA) and gauge unification, can predict 
an LSP consistent with a relic abundance in  the range (\ref{eq:in2}) and
experimentally interesting detection rates.
The space of SUSY parameters compatible with a $CP-$even Higgs mass 
$m_h\approx 114 \ {\rm GeV}$ will restrict the cosmological preferred area
to the region where $\tilde{\chi}-\tilde{\tau}$ coannihilations effects 
are sizeable. 

We will consider the MSSM which described in 
detail in Ref.\cite{cdm} and 
closely follow the notation as 
well as the renormalisation group (RG) and radiative electroweak 
symmetry breaking analysis of this reference.
The full one-loop radiative 
corrections to the effective potential for the electroweak 
symmetry breaking 
and one-loop corrections to 
certain particle masses
which have been evaluated in Ref.\cite{pierce} 
(Appendix E) are incorporated as described in  Ref.\cite{cdm2}.
We do not impose any unification condition to the Yukawa couplings, however,
even though we maintain our calculation
in the large $tan\beta$ regime. We incorporate in our calculation
the two-loop corrections to the CP-even neutral Higgs boson mass matrix by 
using the {\it FeynHiggs} subroutines Ref.\cite{fh}
this improvement is significant, especially for the lighter eigenvalue, at
large values of $\tan\beta$. 

\par
We restrict our a study to the case of 
universal soft SUSY breaking terms at the Grand Unification scale, $M_{GUT}$, 
i.e., a 
common mass for all scalar fields $m_0$, a common gaugino mass 
$M_{1/2}$ and a common trilinear scalar coupling $A_0$. Our effective 
theory below $M_{GUT}$ then 
depends on the parameters ($\mu_0=\mu(M_{GUT})$)
\[
m_0,\ M_{1/2},\ \mu_0,\ A_0,\ \alpha_G,\ M_{GUT},\ h_{t},\ ,\ h_{b},
\ h_{\tau},\ \tan\beta~,  
\]
where $\alpha_G=g_G^2/4\pi$ ($g_G$ being the GUT gauge coupling 
constant) and $h_t, h_b, h_\tau $ are respectively the top, bottom and 
tau Yukawa coupling constants at $M_{GUT}$. The values of $\alpha_G$ and 
$M_{GUT}$ and $|\mu_0|$ are obtained as described in Ref.\cite{cdm}.

\par

For a specified value of $\tan\beta$ at the common SUSY threshold, 
defined as $M_S=\sqrt{m_{\tilde t_1}m_{\tilde t_2}}$, we determine $h_{t}$ at 
$M_{GUT}$ by fixing the top quark mass at the center of its 
experimental range, $m_t(m_t)= 166\ \rm{GeV}$. The value
of  $h_{\tau}$ at $M_{GUT}$ is  fixed by using the running tau lepton
mass at $m_Z$, $m_\tau(m_Z)= 1.746 \ \rm{GeV}$. We also incorporate the
SUSY threshold correction to $m_\tau(M_S)$ from the approximate
formula of Ref.\cite{pierce}.

The value of $h_{b}$ at $M_{GUT}$ is obtained such that 
\beq
m_b(m_Z)_{SM}^{\overline{DR}}=2.90\pm.14 \ {\rm GeV}
\label{mbot}
\eeq
in agreement with the experimental value given in  \cite{mb}. The SUSY 
correction \cite{hall} to the bottom quark mass is 
known to increase with $\tan\beta$. This 
correction originates mainly from squark/gluino and squark/chargino 
loops and has the same  sign as  $\mu$ (in our convention, which is the 
oposite to the one used in  Ref.~\cite{cdm,pierce,cdm2}). With the 
values of the SUSY parameters presented in Fig. 1, the SUSY correction to 
$m_b$ decreases by a 4--5\% as $m_{\tilde{\chi}}$ goes from the lower to the 
higher value in the considered range. Following the analysis of 
\cite{mbsm}, the values of eq.~\ref{mbot} correspond to 
$ m_b(m_b)_{SM}^{\overline{MS}}=4.2\pm.2~{\rm GeV}$ which is in the range
quoted in Ref.~\cite{pdb}.


The choice  $\mu>0$ leads to a constraint on the parameter space coming from
the lower bound on $b\rightarrow s \gamma$. As a result the relatively 
light vlaues for 
$m_{\tilde{\chi}}$ and the obtained detection rates are suppressed. 
From the 
experimental results \cite{cleo} we consider a range within the  
95\% C.L. :
\beq
2.0\times10^{-4} <BR(b\rightarrow s \gamma)<4.5\times10^{-4}
\label{eq:bsg}
\eeq

Our determination of $BR(b\rightarrow s \gamma)$ follows the procedure 
described in ref.~\cite{cdm2}. However appropiate calculations for the 
next-to-leading order (NLO) QCD corrections to the SUSY contribution 
at large values of $\tan\beta$ have  become available \cite{NLO}. 
The SUSY contribution decreases as $M_S$ increases, 
we found that the 
lower bound in (\ref{eq:bsg}) determines a lower limit for $m_{\tilde{\chi}}$.
Therefore in order to perform a consistent prediction for 
$BR(b\rightarrow s \gamma)$ we have implemented the NLO QCD correction 
from the first Ref. \cite{NLO} to the analysis described in~\cite{cdm2}.  
 
The GUT values of $M_{1/2}$ and  $m_{0}$ are traded by the value of the 
mass of the LSP and the mass splitting between the
LSP and the NLSP (next to lightest supersymmetric particle), which in our
approach is one of the $\tilde{\tau}$'s. The parameter
 $\Delta_{\tilde\tau_2}=(m_{\tilde\tau_2}-m_{\tilde\chi})/
m_{\tilde\chi}$ is one of our basic independent parameters. This is 
motivated by the role played by coannihilations $\tilde\chi-\tilde{\tau}$
 in the 
determination of the cosmologically preferred area of the parameter space.
We keep $A_0=0$ in all our calculations.

The composition of the LSP on the model under consideration can be written
in the basis of the gauge and Higgs bosons superpartners as:
\begin{equation}
\tilde{\chi}\equiv\tilde{\chi}^0=C_{11}\tilde{B}+C_{12}\tilde{W}+
C_{13}\tilde{H}_1+C_{14}\tilde{H}_2.
\end{equation}
In the parameter space we study, $\tilde{\chi}$
is mostly a gaugino 
with $P=|C_{11}|^2+|C_{12}|^2 > .9$, with the Bino component being the most 
dominant one.

\begin{figure}
\hspace*{-0.5in}
\begin{minipage}[b]{9in}
\epsfig{figure=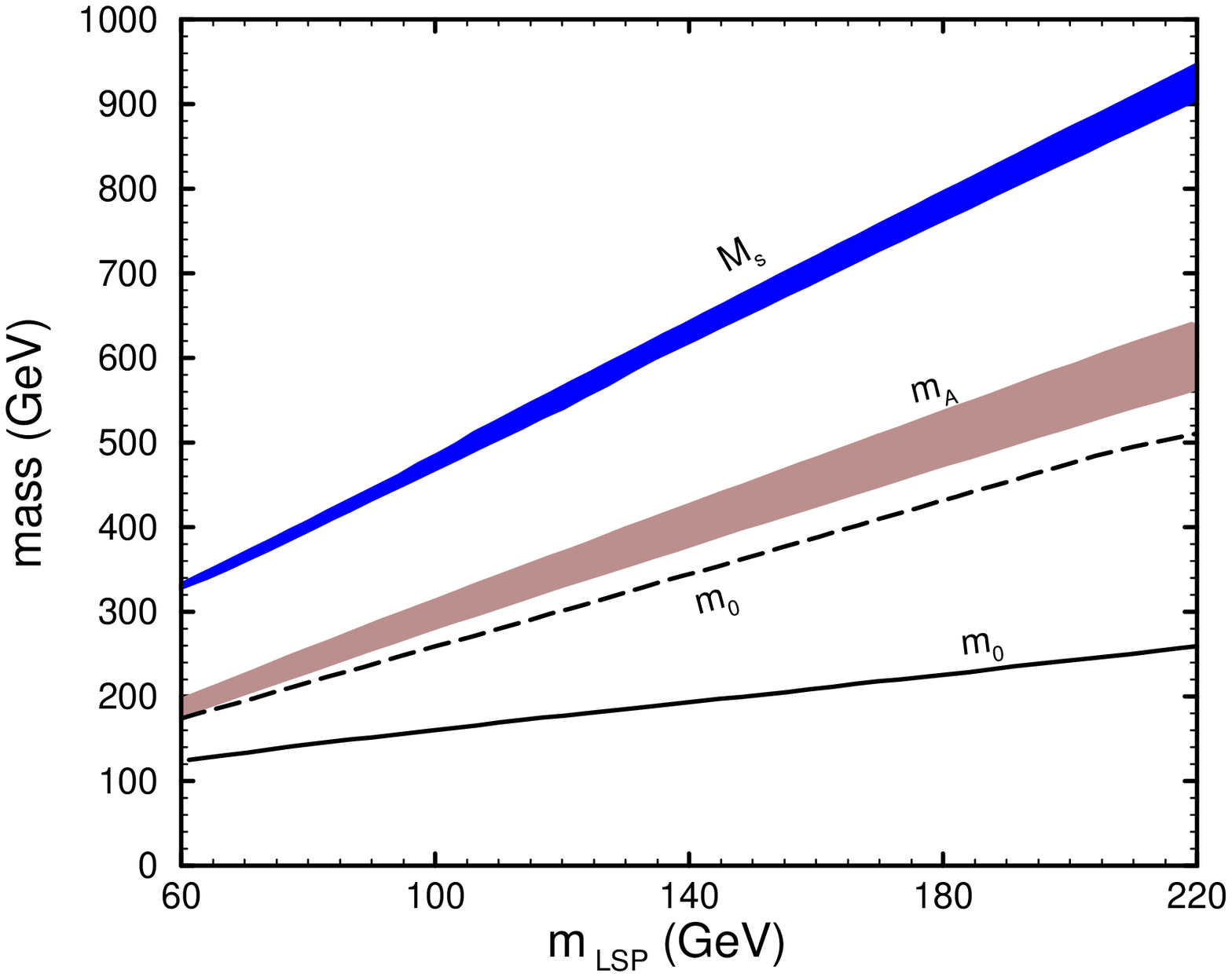,height=3.2in,width=3.2in,angle=0}
\epsfig{figure=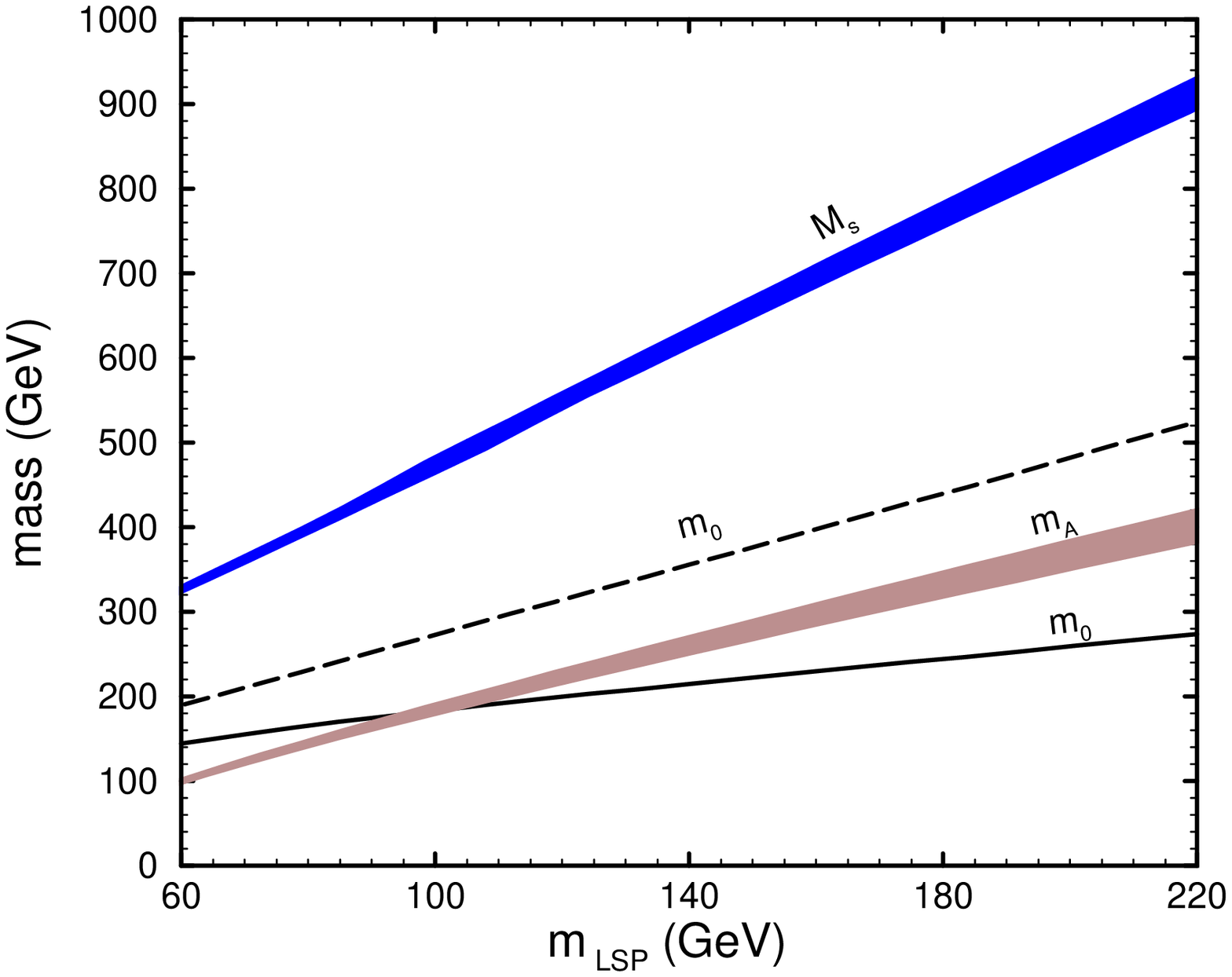,height=3.2in,width=3.2in,angle=0}
\end{minipage}
\medskip
\caption{The values of  $m_0$, $m_{A}$ 
and $M_S$ as functions of $m_{\tilde\chi}$ or $m_{\tilde{\chi}}$ for 
$\tan\beta=40$ (left) and $\tan\beta=52$ (right), 
$\mu>0$, $A_0=0$, the upper boundaries (lower) on the shaded areas
and the upper (lower) dashed line
corresponds to $m_{\tilde\tau_2}= 2 \times m_{\tilde\chi}$ 
($m_{\tilde\tau_2}=m_{\tilde\chi}$).
\label{maslsp}}
\end{figure}

The parameter space considered  relevant for our analysis is displayed in
Fig.\ref{maslsp}.
 For convenience we plot the $M_S$, 
$m_A$ and $m_0$ versus the the LSP mass. Since in our analysis 
the LSP is mostly a bino, one finds  $M_{1/2}\approx 2.5  m_{\tilde{\chi}}$. 
Two values of $\tan\beta$ were considered, one higher and one lower, which we
view as representative. The higher one, 
$\tan\beta=52$, corresponds approximately to the unification of the 
tau and top Yukawa couplings at $M_{GUT}$. We should clarify, however, that  
in making this choice we do not adhere to any particular Yukawa unification 
model.  We find that the parameter space with this 
value of $\tan\beta$ is representative for the purpose of this work . 
The lower value, $tan\beta=40$, results in a  $\sigma^{nucleon}_{scalar}$ 
smaller by one order of magnitude as we will see later. The shaded 
areas correspond to the range of
values taken by $m_A$, $M_S$ as  $\Delta_{\tilde{\tau_2}}$ ranges from
0 to 1. The area associated with $m_0$ for the same range of 
$\Delta_{\tilde{\tau_2}}$ is wider as shown by the dashed and solid lines. 
Lower values of $\tan\beta$ will tend to increase the values for $m_A$. 
This will lead to a decrease of 
the detection rates as we will see later. 


 It can be shown   
 \cite{Verg98},\cite{Verg99},\cite{JDV00}
that the non directional rate can be cast in the form:
\beq
R =  \bar{R}~t(a,Q_{min})] 
          [1 + h(a,Q_{min})cos{\alpha})] 
\label{3.55a}  
\eeq
where $a=[\mu_r(A)~b~ \upsilon _0 \sqrt{2}]^{-1}$ with b the nuclear
size parameter,
 $\mu_r(A)$ is the LSP-nucleus reduced mass  and $\upsilon_0$ the
 parameter characterizing the LSP velocity distribution
assumed here to be Maxwell-Boltzmann. 
The quantity $h$ describes the modulation, i.e. the dependence
of the rate on the Earth's motion, which is of no interest to us in
the present work. $Q_{min}$ is the energy transfer cutoff imposed by
 the detector. The parameter $t$ takes care of the folding with the
LSP velocity distribution and the structure
of the nucleus and it also depends on the LSP mass.
 The most important to us parameter is $\bar{R}$, which carries the
essential dependence on the SUSY parameters, and it can be cast in the form:
\beq
\bar{R} =\frac{\rho (0)}{m_{\tilde{\chi}}} \frac{m}{Am_N} \sqrt{\langle
v^2\rangle } [\bar{\Sigma}_{S}+ \bar{\Sigma} _{spin} + 
\frac{\langle \upsilon ^2 \rangle}{c^2} \bar{\Sigma} _{V}]
\label{3.39b}  
\eeq
 Where
 $\rho (0)$ is the LSP density in our vicinity, which will be taken to
be $0.3\ GeV/cm^3$, and m is the detector mass. 
$\bar{\Sigma} _{i}, i=S,V,spin$ are associated with the scalar,
vector and spin  contributions. In the present paper we will consider
only the scalar  contribution, which offers the best chance for measurable
detection rates, in particular for medium-heavy nuclear targets,
 like $^{127}I$. In this case 
\begin{equation}
\bar{\Sigma} _{S} = \sigma_0 ~\mu^2_r(A)  \,
 \{ A^2 \, [ (f^0_S - f^1_S \frac{A-2 Z}{A})^2 \, ] \simeq \sigma^S_{p,\chi^0}
        A^2 (\frac{\mu_r(A)}{\mu _r(N)})^2 
\label{2.9}
\end{equation}
where $m_N$ is the nucleon mass,
 $\mu_r(N)$ is the LSP-nucleon reduced mass,
$\sigma^S_{p,\chi^0}$ is the scalar LSP-proton cross-section and 
\begin{equation}
\sigma_0 = \frac{1}{2\pi} (G_F m_N)^2 \simeq 0.77 \times 10^{-38}cm^2 
\label{2.7} 
\end{equation}

The coherent scattering can be mediated via the the neutral
intermediate Higgs particles (h and H), which survive as physical 
particles. It can also be mediated via s-quarks, via the mixing
of the isodoublet and isosinlet s-quarks of the same charge. In our
model we find that the Higgs contribution becomes dominant and, as a
matter of fact, the heavy Higgs H is more important (the Higgs particle
$A$ couples in a pseudoscalar way, which does not lead to coherence).
As it has been pointed out in Ref.~\cite{bot} large values of  
$\tan\beta$ and low values for the mass of the CP-even Higgs masses
($m_h$ and $m_H$) enhance the Higgs exchange amplitude. This is of 
particular interest in our analysis, since, as $\tan\beta$ increases, the 
electroweak symmetry breaking imposes lower values for for the 
pseudoscalar Higgs mass $m_A$ (see Fig. 1). This implies a lower value 
for of $m_H$. The changes on   $m_h$ are not so important, 
since its value can only move below an upper bound of 
about 120-130 GeV (see for e.g \cite{fh}).
Since the values predicted for the $\mu$-parameter also decrease, the Higgsino
component of the LSP enhances the values of $\sigma^{nucleon}_{scalar}$.
For all the values of the $m_{\tilde{\chi}}$, however, reported 
in the present work the condition $ P>.9$ is maintained. It becomes even more
stringent, $P >.95$, for $m_{\tilde{\chi}}>100 \ \rm{GeV}$.

It is well known that all quark flavors contribute, see e.g. Drees
{\it et al}\cite{Dree}, since the relevant couplings are proportional to
the quark masses.
One encounters in the nucleon not only the usual 
sea quarks ($u {\bar u}, d {\bar d}$ and $s {\bar s}$) but the 
heavier quarks $c,b,t$ which couple to the nucleon via two gluon 
exchange, see e.g. Drees {\it et al}  ~\cite{Dree00} and references
therein.
 
As a result  one obtains an effective scalar Higgs-nucleon
coupling by using  effective quark masses as follows
\begin{center}
$m_u \ra f_u m_N, \ \ m_d \ra f_d m_N. \ \ \  m_s \ra f_s m_N$   
\end{center}
\begin{center}
$m_Q \ra f_Q m_N, \ \ (heavy\ \  quarks \ \ c,b,t)$   
\end{center}
where $m_N$ is the nucleon mass. The isovector contribution is now
negligible. The parameters $f_q,~q=u,d,s$ can be obtained by chiral
symmetry breaking 
terms in relation to phase shift and dispersion analysis.
Following Cheng and Cheng ~\cite{Chen} we obtain
\begin{center}
$ f_u = 0.021, \quad f_d = 0.037, \quad  f_s = 0.140$ 
\quad  \quad  (model B)   
\end{center}
\begin{center}
$ f_u = 0.023, \quad f_d = 0.034, \quad  f_s = 0.400$ 
\quad  \quad  (model C)   
\end{center}
 We see that in both models the s-quark is dominant.
Then to leading order via quark loops and gluon exchange with the
nucleon one finds:
\begin{center}
\quad $f_Q= 2/27(1-\quad \sum_q f_q)$   
\end{center}
This yields:
\begin{center}
\quad $ f_Q = 0.060$    (model B),   
\quad $ f_Q = 0.040$    (model C)   
\end{center}
 There is a correction to the above parameters coming from loops
involving s-quarks \cite {Dree00} and due to QCD effects. 
 Thus for large $tan \beta$ we find \cite {JDV00}:
\begin{center}
\quad $f_{c}=0.060 \times 1.068=0.064,
       f_{t}=0.060 \times 2.048=0.123,
       f_{b}=0.060 \times 1.174=0.070$  \quad (model B)
\end{center}
\begin{center}
\quad $f_{c}=0.040 \times 1.068=0.043,
       f_{t}=0.040 \times 2.048=0.082,
       f_{b}=0.040 \times 1.174=0.047$  \quad (model B)
\end{center}
For a more detailed discussion we refer the reader to 
Refs~\cite{Dree,Dree00}.
 With the above ingredients and employing the obtained 
constraints on the SUSY parameter
space one can proceed to calculate first the parameter $f_S^0$ associated with
the scalar contribution.
 Then combine it with the effective Higgs-quark couplings $f_q$ and $f_Q$ to
obtain the nucleon cross section 
(the contribution to the scalar coupling arising from the
L-R s-quark mixing is in our model negligible). 

\begin{figure}
\epsfig{figure=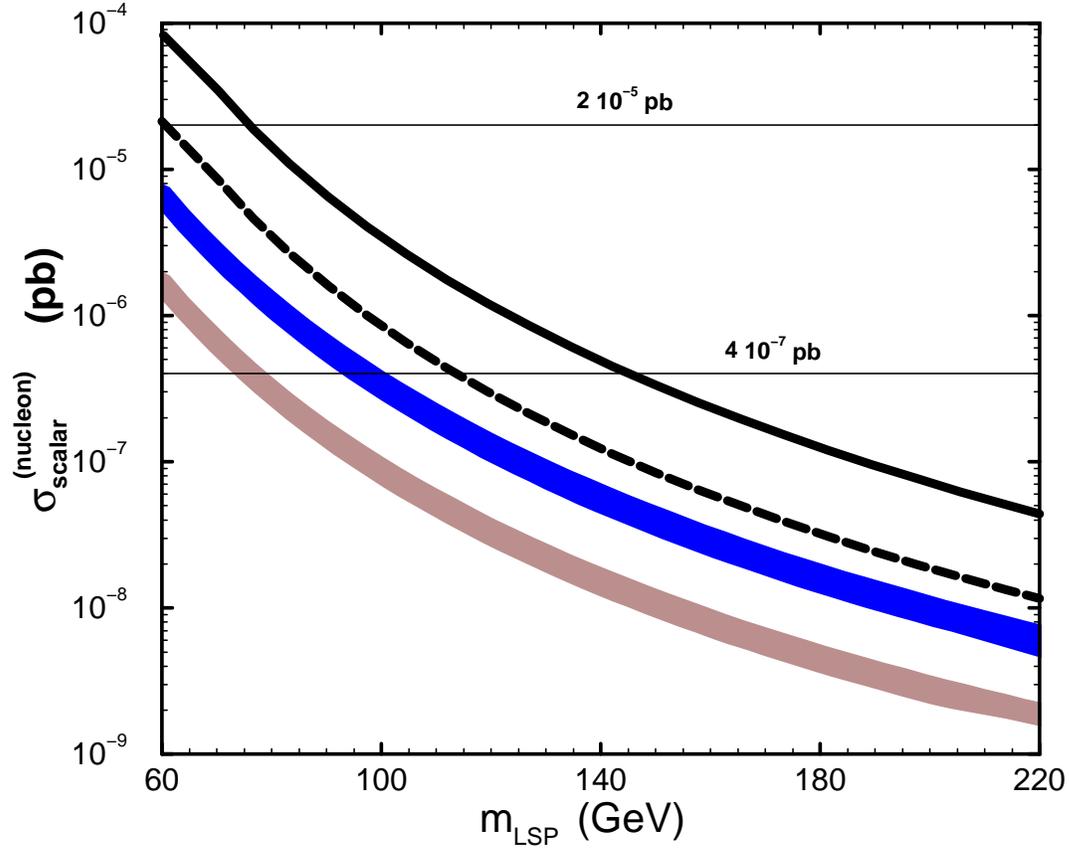,height=4.5in,angle=0}
\medskip
\caption{Values for $\sigma_{scalar}^{(nucleon)}$, corresponding to the input
parameters presented in Fig. \ref{maslsp}. The upper lines corresponds to
$\tan\beta=52$, the solid (dashed) one was obtained using model C (B). 
The effect of changing 
with $\Delta_{\tilde\tau_2}$ from 0 to 1 is  not sizeable for $\tan\beta=52$.
The bands  are obtained using $\tan\beta=40$, the darker shaded band 
corresponds to model C with the upper bound corresponding 
to $\Delta_{\tilde\tau_2}=1$ and the lower to  
$\Delta_{\tilde\tau_2}=0$. The light shaded band corresponds to model
B, for the same range of values of $\Delta_{\tilde\tau_2}$
\label{siglsp}}
\end{figure}

The values for $\sigma^{nucleon}_{scalar}$ for the two nucleon models 
described above and the input SUSY parameters presented in Fig.1 
are shown in Fig. \ref{siglsp}. 
 As we can see nucleon model C leads to higher cross sections.
For $\tan\beta=52$ the effect of changing  $\Delta_{\tilde\tau_2}$ 
(equivalently $m_0$ in the range shown in Fig.1)
from 0 to 1 has a small effect on the cross section 
(not sizeable in the figure) while for $\tan\beta=40$ the lines are widened 
and become 
bands. This effect can be understood in the light of Fig. 1. As $\tan\beta$ 
decreases the values of $m_A$ (and hence $m_H$ and the $\mu$-parameter) 
increase and span a larger area as $\Delta_{\tilde\tau_2}$ varies from 
0  to 1. This leads to a substantial decrease in the cross section 
for the lower value of $\tan\beta$.
 

The fact that ($\tilde\chi$) is mostly a 
$\tilde B$ implies that the main contribution 
to its annihilation cross section arises from s-fermion 
(squark, s-lepton) exchange in the t- and u-channel 
leading to $f\bar f$ final states ($f$ is a quark 
or lepton). If, however, the mass of $\tilde\chi$ is close to the one 
of the NLSP, coannihilations between the two particles must be 
taken into account \cite{coan}. The inclusion 
these coannihilation effects results in a dramatic reduction of the 
($\tilde\chi$) relic abundance as the two lightest SUSY particles
approach in mass \cite{cdm,drees,ellis}. 
We estimate the  relic abundance of the LSP ($\tilde\chi$), 
by employing 
the analysis of Ref.\cite{cdm} which is appropriate for 
large $\tan\beta$ and includes coannihilations $\tilde\chi-\tilde{\tau}$,
suitable for Bino like LSP. To the list of coannihilation channels of 
presented in \cite{cdm}, one should add the t-channed sneutrino 
exchange in the $W^+W^-$ and  $H^+H^-$ production.

We should, at this point, clarify that 
in the parameter space considered
here no resonances in the s--channels were found. In other words
the s--channel exchange of 
A , h, H, Z into  $\tilde{\tau_2} \tilde{\tau_2}^*$ never becomes
resonant in the parameter space of our analysis. We can see in Fig. 1, however,
that a line $mass= 2 m_{\tilde{\chi}}$ will be above of the $m_A$ region 
for the 
case of $\tan\beta=52$, while for $\tan\beta=40$ it will not. 
However we should emphasize here, that the position the $m_A$ band 
displayed is Fig.1 respect a line of $mass= 2 m_{\tilde{\chi}}$ is 
very sensitive to small changes in $\tan\beta$ and  
the values $m_t$,$m_b$ and the GUT values for $A_0$ and $m_0$. 
Therefore, at the large values of $\tan\beta$ it is possible 
to find sectors of the space of parameters where 
$m_A\approx m_{\tilde{\chi}}$, in these cases the the adequate treatment
of the Higgs mediated anihilation channels will be determining for an accurate
calculation of $\Omega_{LSP}~h^2$. 
The choice of parameter space in the two examples we present is aimed
to illustrate the decisive role of $\tan\beta$ in the LSP 
detection rates. In the two scenarios we choose, anihilation resonant 
channels are not present and coannihilations are required in order 
to predict a cosmologically desirable LSP relic abundance. 

\begin{figure}
\hspace*{-0.5in}
\begin{minipage}[b]{9in}
\epsfig{figure=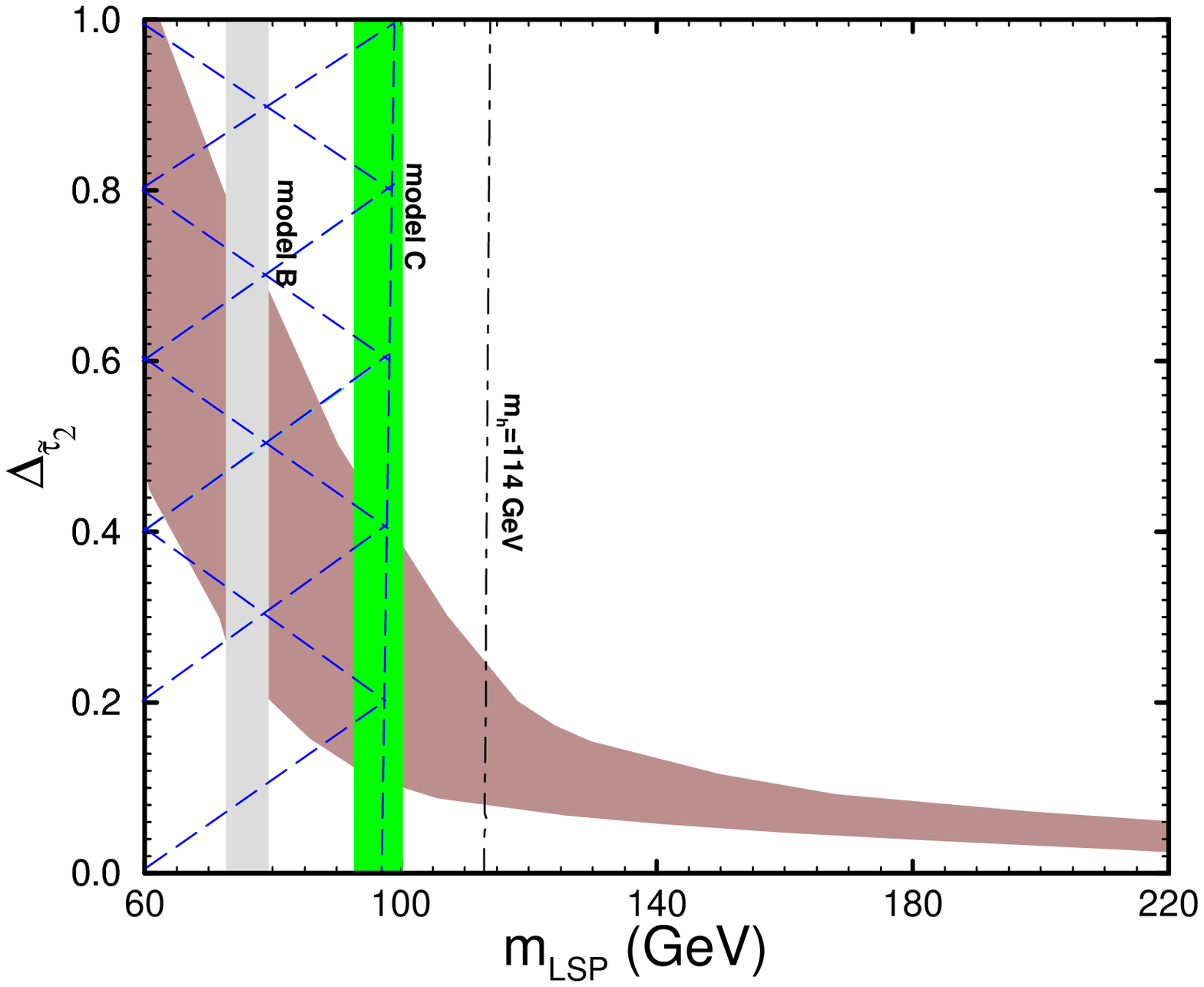,height=3.2in,width=3.2in,angle=0}
\epsfig{figure=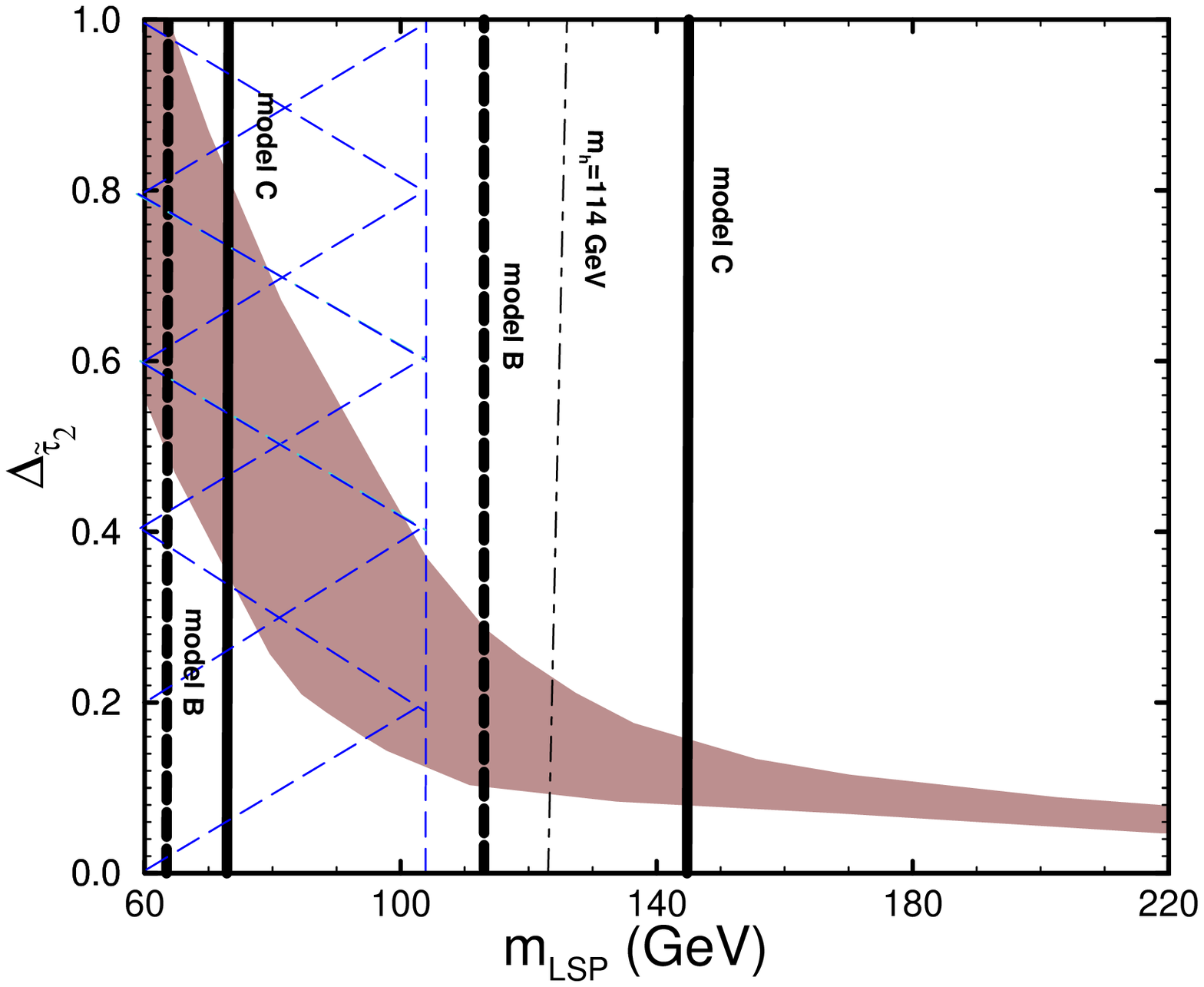,height=3.2in,width=3.2in,angle=0}
\end{minipage}
\medskip
\caption{The cosmologically allowed region in the 
$m_{\tilde{\chi}}-\Delta_{\tilde\tau_2}$ plane (shaded area) for $\tan\beta=40$ (left)
and $\tan\beta=52$ (right). 
The vertical bands in graph on  the left
correspond to the bound $\sigma_{scalar}^{(nucleon)}=4 \cdot10^{-7} pb$, 
obtained
from figure 2 for both models.
The vertical lines on the graph of the right 
correspond to the bounds $\sigma_{scalar}^{(nucleon)}=4 \cdot10^{-7} pb$
(lines towards the right of the graph) and $2 \cdot 10^{-5} pb$ for 
the models indicated. The marked areas on the left are excluded by 
$b\rightarrow s \gamma$.
\label{area50}}
\end{figure}

Fig.~\ref{area50} shows the cosmologically allowed area from CDM 
considerations corresponding to the parameter space described in Fig.1. 
The $b\rightarrow s \gamma$ constraint will exclude the parameter 
space marked on the left of the figures. 
The larger
range on the the values of the LSP masses corresponds to the narrow
band of the mass splitting between the LSP--NLSP, which enhances the effect
of coannihilations. We find the lower of the bounds of Eq. (\ref{eq:in2})
in the displayed area corresponds $\Delta_{\tilde\tau_2}=.045$ for 
$\tan\beta=52$ and  $\Delta_{\tilde\tau_2}=.03$ for 
$\tan\beta=40$, the effect of the $\tilde{\chi}-\tilde{\tau}$ coannihilation
disappears for $\Delta_{\tilde\tau_2}\approx.25$. 
Higher values for 
$m_{\tilde\tau_2}$, enhance the neutralino relic abundance. We
found that for values of $m_h>105 \rm{GeV}$, a choice of
parameters such that  $m_{\tilde\tau_2} > 2 m_{\tilde\chi}$
leads to values for $\Omega_{LSP} h^2 >.22$, the upper limit
we consider as cosmologically desirable.

The $b\rightarrow s \gamma$ constraint is more restictive than the
bounds on the masses of the  $\tilde\tau$ and chargino.
Lines of constant mass for lighter $\tilde\tau$ can be 
easily deduced from the coordinates of the graphs in Fig.~\ref{area50}, 
the ones corresponding to values
of about 95 GeV will start intersecting the cosmologically favored area
in both examples, they would lay in the area excluded by the 
$b\rightarrow s \gamma$ constraint. Lines of constant chargino mass 
would  appear in the 
graphs of Fig.~\ref{area50} as almost vertical lines. The  line 
corresponding to 
$m_{\tilde{\chi}^+}=110\ \rm{GeV}$ will be located at 
$m_{\tilde{\chi}}=64\ \rm{GeV}$ for $\tan\beta=52$ and 
$m_{\tilde{\chi}}=62\ \rm{GeV}$ for $\tan\beta=40$. In the space of 
parameters under
consideration, the difference 
$m_{\tilde{\chi}^+}-m_{\tilde{\chi}}$ is large enough to consider 
any effect of  $\tilde{\chi}^+-\tilde{\chi}$
coannihilations in the $\tilde{\chi}$ relic abundance. Other 
scenarios allowing sizeable  
$\tilde{\chi}^+-\tilde{\chi}$ coannihilation effects
may be possible for large $\tan\beta$. Such a study, however, is beyond the 
scope of the present work.

We find that the possible values of lightest neutral CP 
even Higgs mass, imposes a severe constraint on the cosmologically 
allowed area. We also found that, in a fashion similar to the results 
presented  in Ref.~\cite{egno} for the low $\tan\beta$ scenario, 
 in the large $\tan\beta$ case a value of 
$m_h=114\ \rm{GeV}$ will push the cosmologically allowed area to the 
values of the parameters where $\tilde{\chi}-\tilde{\tau}$ 
coannihilations are important (see  Fig.~\ref{area50}).
Changes in the values of  $\Delta_{\tilde\tau_2}$
will not affect significatively the values of $m_h$ since, as we can see in 
the shaded areas of Fig. \ref{maslsp}, the changes in 
$\Delta_{\tilde\tau_2}$ imply small changes on the values
of $m_A$ and $M_S$. Conversely small changes on $m_h$ are translated
in significative displacements of the line of constant $m_h$  shown 
in Fig.~\ref{area50}. Indeed this line would  be better 
understood as a band
of approximately 20 GeV width on $m_{\tilde{\chi}}$  
for changes on $m_h$ from 113 GeV to 115 GeV (these changes 
can be understood from the shape of the  curves shown, e.g, on the first 
reference of Ref. \cite{fh}). We can note, by observing Fig.  \ref{maslsp}, 
 that under the 
conditions imposed on the parameter space  larger values for $\tan\beta$ 
imply lower values for $m_A$ at the same range of $m_{\tilde{\chi}}$. This 
helps to understand the different position of the lines of constant 
$m_h$ in the two graphs of Fig.~\ref{area50}. 

As we can see in Fig.~\ref{area50}, a value of
 $m_h=114\ \rm{GeV}$ will exclude the the space of parameters suitable 
for detection for $\tan\beta=40$ while for $\tan\beta=52$, the line 
corresponding to the same value for $m_h$ lies inside the range 
of Eq. (\ref{eq:in3}) for model C. The lines corresponding
to $m_h=105\ \rm{GeV}$ would be also almost vertical lines located at the 
abscissa  $m_{\tilde{\chi}}\approx \ 63\rm{GeV}$ for $\tan\beta=52$ and below 
the displayed area for $\tan\beta=40$. Therefore the dependence of our 
conclusions on $m_h$ is clear.

\begin{figure}
\begin{minipage}[b]{8in}
\epsfig{figure=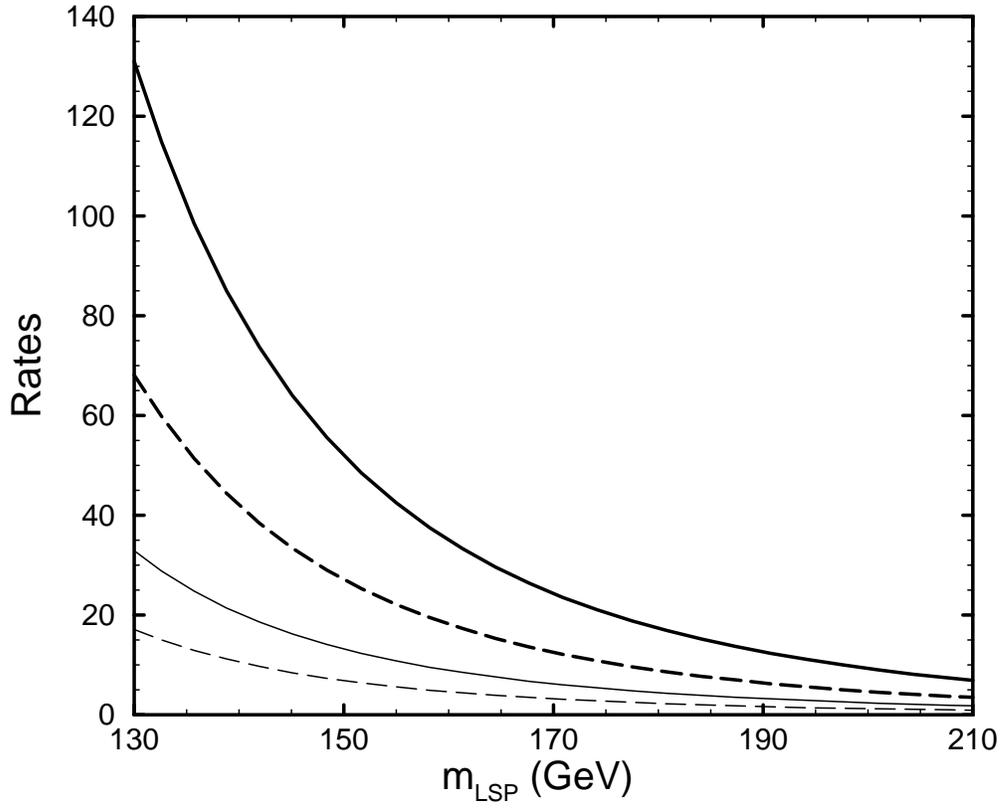,height=4.3in,angle=0}
\end{minipage}
\medskip
\caption{The Total detection rate per $(kg-target)yr$ vs the LSP mass
in GeV  in the case of 
$^{127}I$, corresponding to  model B (thick lines) and Model C (fine 
lines). We used the parameter space corresponding to 
$\tan\beta=52$ and $m_h >$ 114 GeV.
On the solid curves we used no detector threshold energy cut off.
On the dashed ones, the value $10~KeV$ was employed. 
\label{rate}}
\end{figure}

 With the above ingredients one can proceed, as outlined above, with the 
calculation of the event rates.  For a typical
nucleon cross section, taken, e.g.,  from the set  associated with 
$\tan\beta=52$, exhibited in Fig. \ref {siglsp}, 
we plot in Fig. \ref{rate} the event rate as a function of the LSP 
mass in the case of the popular target $^{127}I $. Since the detector 
cut off energy is not so well known, we consider two possibilities, namely 
$Q_{min}=0$ and $10~KeV$. The values of the quantity $t$,
 see Eq. (\ref {3.55a}),
employed here have been taken from  our earlier work \cite{Verg00}. In these 
plots we have excluded LSP masses less than $130~GeV$
to avoid a logaritmic scale. They can easily be extrapolated to values of 
$m_{LSP}$ down to 100 $GeV$. Anyway event rates in this LSP mass region are 
very large in our model and it unlikely that they would have been missed by 
the existing experiments.

In summary, we have found that the most popular version of the MSSM with 
gauge unification and universal boundary conditions at the GUT scale,
and a parameter space determined by large values of $\tan\beta$,
can accommodate a cosmologically suitable LSP relic abundance and
predict detection rates, which can be tested in current or projected
experiments.

We should mention, however, that 
 the calculated detection rates can vary by orders of magnitude, depending
on the yet unknown LSP mass. We will
not be concerned here with uncertainties, which are essentially of 
experimental nature, like the threshold energy cut off 
imposed by the detector. This 
reduces the rates, especially for small LSP masses. In the LSP mass range 
considered here the reduction is about a factor of two (see Fig. \ref {rate}). 
The next uncertainty comes from estimating the heavy quark contribution in the
nucleon cross section. This seems to be under control. We take the difference
between the models B and C discussed above as an indication of such
 uncertainties. They seem to imply uncertainties no more than factors of two 
(see Fig. \ref{siglsp}). 
The nuclear uncertainties for the coherent process are even smaller. 

 We believe, therefore, that, concerning the direct LSP detection event rates 
the main uncertainties come from the fact that the SUSY parameter space 
is not yet sufficiently constrained. The parameter space may be sharpened
by the accelerator experiments, even if the LSP is not found.  We should
mention here, in particular, the Higgs searches, since, as we have seen, the 
role of the Higgs particles in direct SUSY dark matter detection is crucial. 
It is not an exaggeration to say that
the underground and accelarator experiments are complementary and should
achieve a symbiosis.

\par
Note added:
After the first version of this work was completed the Muon $(g-2)$ 
Collaboration had published a new meassurement of the anomalous 
magnetic moment of the muon \cite{g2}. The result presented differs 
from the SM prediction by 2.6$\sigma$. If this discrepancy is 
attributed to SUSY contributions, the $(g-2)$ may impose a further
constraint on the parameter space presented here. When we consider
the reported discrepancy at the $2-\sigma$ level as the allowed 
range for the SUSY contribution, the values we found do not have 
a significative impact in the paramter space considered in this work.
Using the calcualtion presented in ref.\cite{his} the upper bound will
exclude mostly the area already excluded by $b\rightarrow s \gamma$, 
while the area excluded by the lower bound corresponds to values of 
$m_{\tilde{\chi}}>300 \rm{GeV}$. We are thus in agreement with Ref.\cite{ENO} 
in the space of parameters we overlap.

We would like to thank S. Khalil for useful discussions and 
for reading the manuscript and T. Falk for pointing out the 
missing coannihilation channels on the list of Ref.~\cite{cdm}.
M.E.G. thanks the University of Ioannina for kind hospitality.
This work was supported by the European Union under TMR contract 
No. ERBFMRX--CT96--0090 and $\Pi E N E \Delta~95$ of the Greek 
Secretariat for Research.  

\def\ijmp#1#2#3{{ Int. Jour. Mod. Phys. }{\bf #1~}(#2)~#3}
\def\pl#1#2#3{{ Phys. Lett. }{\bf B#1~}(#2)~#3}
\def\zp#1#2#3{{ Z. Phys. }{\bf C#1~}(#2)~#3}
\def\prl#1#2#3{{ Phys. Rev. Lett. }{\bf #1~}(#2)~#3}
\def\rmp#1#2#3{{ Rev. Mod. Phys. }{\bf #1~}(#2)~#3}
\def\prep#1#2#3{{ Phys. Rep. }{\bf #1~}(#2)~#3}
\def\pr#1#2#3{{ Phys. Rev. }{\bf D#1~}(#2)~#3}
\def\np#1#2#3{{ Nucl. Phys. }{\bf B#1~}(#2)~#3}
\def\npps#1#2#3{{ Nucl. Phys. (Proc. Sup.) }{\bf B#1~}(#2)~#3}
\def\mpl#1#2#3{{ Mod. Phys. Lett. }{\bf #1~}(#2)~#3}
\def\arnps#1#2#3{{ Annu. Rev. Nucl. Part. Sci. }{\bf
#1~}(#2)~#3}
\def\sjnp#1#2#3{{ Sov. J. Nucl. Phys. }{\bf #1~}(#2)~#3}
\def\jetp#1#2#3{{ JETP Lett. }{\bf #1~}(#2)~#3}
\def\app#1#2#3{{ Acta Phys. Polon. }{\bf #1~}(#2)~#3}
\def\rnc#1#2#3{{ Riv. Nuovo Cim. }{\bf #1~}(#2)~#3}
\def\ap#1#2#3{{ Ann. Phys. }{\bf #1~}(#2)~#3}
\def\ptp#1#2#3{{ Prog. Theor. Phys. }{\bf #1~}(#2)~#3}
\def\plb#1#2#3{{ Phys. Lett. }{\bf#1B~}(#2)~#3}
\def\apjl#1#2#3{{ Astrophys. J. Lett. }{\bf #1~}(#2)~#3}
\def\n#1#2#3{{ Nature }{\bf #1~}(#2)~#3}
\def\apj#1#2#3{{ Astrophys. Journal }{\bf #1~}(#2)~#3}
\def\anj#1#2#3{{ Astron. J. }{\bf #1~}(#2)~#3}
\def\mnras#1#2#3{{ MNRAS }{\bf #1~}(#2)~#3}
\def\grg#1#2#3{{ Gen. Rel. Grav. }{\bf #1~}(#2)~#3}
\def\s#1#2#3{{ Science }{\bf #1~}(19#2)~#3}
\def\baas#1#2#3{{ Bull. Am. Astron. Soc. }{\bf #1~}(#2)~#3}
\def\ibid#1#2#3{{ ibid. }{\bf #1~}(19#2)~#3}
\def\cpc#1#2#3{{ Comput. Phys. Commun. }{\bf #1~}(#2)~#3}
\def\astp#1#2#3{{ Astropart. Phys. }{\bf #1~}(#2)~#3}
\def\epj#1#2#3{{ Eur. Phys. J. }{\bf C#1~}(#2)~#3}

\newpage
\pagestyle{empty}


\end{document}